# Shielded Ohmic Contact (ShOC) Rectifier: A New Metal-Semiconductor Device with Excellent Forward and Reverse Characteristics

*M. Jagadesh Kumar*

Department of Electrical Engineering,

Indian Institute of Technology, Delhi,

Hauz Khas, New Delhi – 110 016, INDIA.

Email: mamidala@ieee.org Fax: 91-11-2658 1264

**Abstract**

We report a new structure, called the Shielded Ohmic Contact (ShOC) rectifier which utilizes trenches filled with a high barrier metal to shield an Ohmic contact during the reverse bias. When the device is forward biased, the Ohmic contact conducts with a low forward drop. However, when reverse biased, the Ohmic contact is completely shielded by the high barrier Schottky contact resulting in a low reverse leakage current. Two dimensional numerical simulation is used to evaluate and explain the superior performance of the proposed ShOC rectifier.

# Introduction

Applications requiring low power dissipation and fast switching speed widely use low barrier Schottky rectifiers. One has to always make a trade-off between the forward voltage drop and the reverse leakage current, since a low barrier metal gives small forward voltage drop but at the cost of high leakage current and the high barrier metal gives a lower leakage current but with a high forward voltage drop. To overcome the above problem, several novel lateral Schottky rectifiers have been reported in the recent past [1-5]. Among them, in the lateral merged double Schottky (LMDS) rectifier [1,2], two Schottky barriers are used - the low barrier Schottky contact conducts during forward bias and the high barrier Schottky contact effectively shields the low barrier Schottky contact during the reverse bias. One drawback of the LMDS rectifier is that it requires two Schottky metals. In this paper, we report a new device called the Shielded Ohmic Contact (ShOC) rectifier in which the low barrier Schottky contact in the LMDS rectifier is replaced by an Ohmic contact so that during the forward bias, the Ohmic contact conducts and is effectively shielded by the high barrier Schottky contact during the reverse bias. Using numerical simulations we demonstrate that using the ShOC rectifier, which uses only one high barrier Schottky contact, we can still realize low forward drop, low reverse leakage current and high breakdown voltage similar to that of the LMDS rectifier which uses two Schottky contacts.

## Simulation results and discussion

A cross-sectional view of the ShOC rectifier is shown in Fig.1 which is implemented in MEDICI[6], a 2D device simulator. The tunneling Ohmic contact formed by the high barrier metal on the diffused $N^+$ region between the two high-barrier metal trenches forms the anode contact of the device. This $N^+$ region can be formed before creating the trenches in the oxide window region. The cathode contact is placed symmetrically on both sides of the anode. To reduce the electric field crowding at the trench edges, metal field termination is used[7]. We have compared the ShOC rectifier with LMDS and the conventional low barrier Schottky (LBS) and high barrier Schottky (HBS) structures. The parameters used for the simulation of these rectifiers are given in Table 1.

The simulated current voltage characteristics of the ShOC rectifiers are compared with that of the LMDS, conventional low barrier and high barrier Schottky structures in Fig. 2. We can see that the forward characteristics (as shown in Fig. 2(a)) and the reverse characteristics (as shown in Fig. 2(b)) are approximately identical for the ShOC and LMDS rectifiers. Just as in the case of the LMDS rectifier, the ShOC rectifier also exhibits a higher breakdown voltage when compared to the LBS and HBS rectifiers. The improved performance of the ShOC rectifier can be understood from the current flow lines shown in Fig.3. When the ShOC rectifier is forward biased, the Ohmic contact conducts all the forward current giving rise to a low forward voltage drop. However, under reverse bias conditions, we notice that the Ohmic region is effectively shielded by the high barrier trench and the reverse leakage current corresponds to that of the high-barrier trench. In the above simulations, we have taken an optimum value for the Ohmic contact width ($m$=0.2 μm) at the given trench depth ($d$=1.5 μm). Although we have chosen a trench aspect ratio($d/w$) equal to 7.5, our simulations show that the trench width $w$ can be increased to reduce the trench aspect ratio without affecting the device characteristics.

The performance of the ShOC rectifier, i.e. the how efficiently the Ohmic region can be shielded by the high-barrier trench is, however, dependent on the $d/m$ ratio. The forward voltage drop (at a forward current density of 100 A/cm$^2$) and the reverse current density (at a reverse voltage of 25V) versus the *d/m* ratio are shown in Fig. 4 ( with *d* fixed at 1.5 μm and *m* as a variable) and in Fig. 5 ( with *m* fixed at 0.2 μm and *d* as a variable) We can easily observe from Fig. 4 that as *m* increases for a fixed *d*, the shielding provided by the high barrier trench becomes inefficient and the Ohmic contact region plays a greater role in current conduction. This will result in a reduced forward voltage drop but will also increase the reverse leakage current. Therefore, the *d/m* ratio should be chosen such that the reverse leakage current is at its minimum. Both the forward voltage drop and reverse leakage current will approximately match with that of the LMDS for a *d/m* ratio of 7 or above. On the other hand, we observe from Fig. 5 that for a fixed *m*, if *d* is decreased, the Ohmic contact region is again not efficiently shielded for a *d/m* ratio less than 7. The important observation that needs to be made here is that the Ohmic contact region is not efficiently shielded if *m* is long in relation to *d*. Therefore, for the given trench depth in our simulations, we have chosen the Ohmic contact width *m* to be 0.2 μm so that *d/m* ratio is greater than 7. Thus for a given trench depth, by choosing an appropriate width for the Ohmic region, we can easily realize the same current voltage performance for the ShOC rectifier as compared to the LMDS rectifier.

## Conclusion

A novel shielded ohmic contact rectifier (ShOC) having a high barrier metal filled in a trench surrounding an Ohmic contact is presented. The results show that this structure is as good as the previously reported LMDS structure with the advantage of using only one high barrier Schottky metal contact. The ShOC rectifier also exhibits significantly enhanced breakdown voltage compared to the conventional high barrier or low barrier Schottky rectifiers. The combined low forward voltage drop and excellent reverse blocking capability make the proposed ShOC rectifier attractive for use in low-loss high speed smart power IC applications.

**Table 1. MEDICI parameters for the LBS, HBS, ShOC and LMDS structures**

| Parameter | Value |
| --- | --- |
| $N^+$ doping for Ohmic contact | $10^{19}$cm$^{-3}$ |
| Drift region doping, $N_D$ | $10^{16}$cm$^{-3}$ |
| Drift region length, L | 12 μm |
| Drift region thickness, $t_{si}$ | 2 μm |
| Buried oxide thickness, $t_{box}$ | 3 μm |
| Field oxide thickness, $t_{ox}$ | 0.5 μm |
| Trench depth, d | 1.5 μm |
| Trench width, w | 0.2 μm |
| Mesa width, m | 0.2 μm |
| Low Schottky barrier height(Ni), $\phi_L$ | 0.57eV |
| High Schottky barrier height(W), $\phi_H$ | 0.67eV |

**Figure captions**

Fig. 1 Cross-section of the Shielded Ohmic Contact (ShOC) rectifier.

Fig. 2 (a) Forward characteristics and (b) Reverse characteristics of the ShOC rectifier compared with that of the LMDS, conventional low-barrier Schottky (LBS) and high-barrier Schottky (HBS) rectifiers

Fig. 3 (a) Forward current flow lines of ShOC rectifier at a forward voltage of 0.3 V and (b) Reverse current flow lines of ShOC rectifier at a reverse voltage of 120 V.

Fig. 4 (a) Forward voltage drop versus *d/m* ratio at a forward current density of 100A/cm$^2$ and (b) Reverse leakage current density versus *d/m* ratio at a reverse bias of 25 V for the ShOC and the LMDS rectifiers. Here *m* is varied for a fixed trench depth *w*.

Fig. 5 (a) Forward voltage drop versus *d/m* ratio at a forward current density of 100 A/cm$^2$ and (b) Reverse leakage current density versus *d/m* ratio at a reverse bias of 25 V for the ShOC and the LMDS rectifiers. Here the trench width *w* is varied for a fixed *m*.

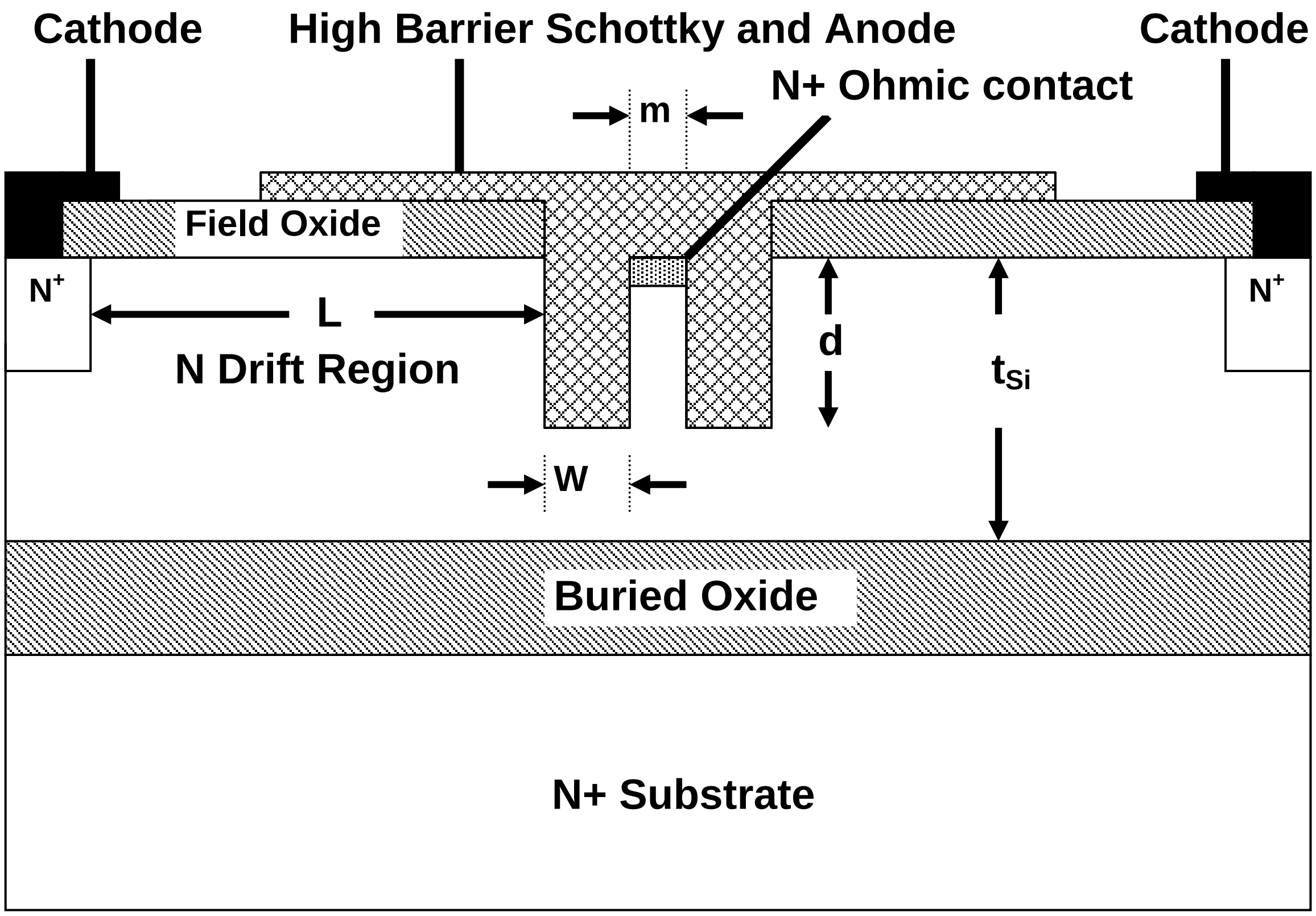

Fig.1

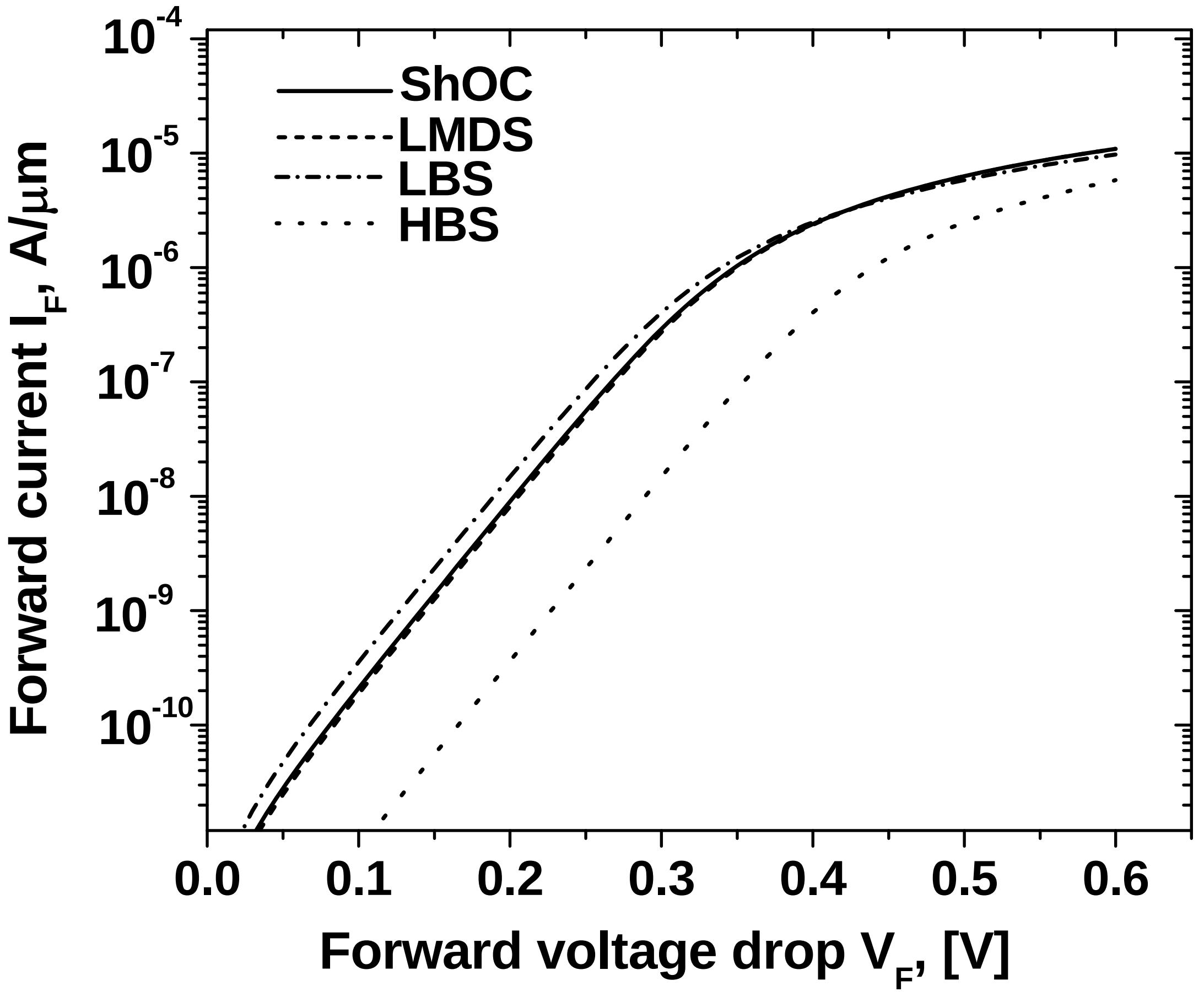

Fig. 2(a)

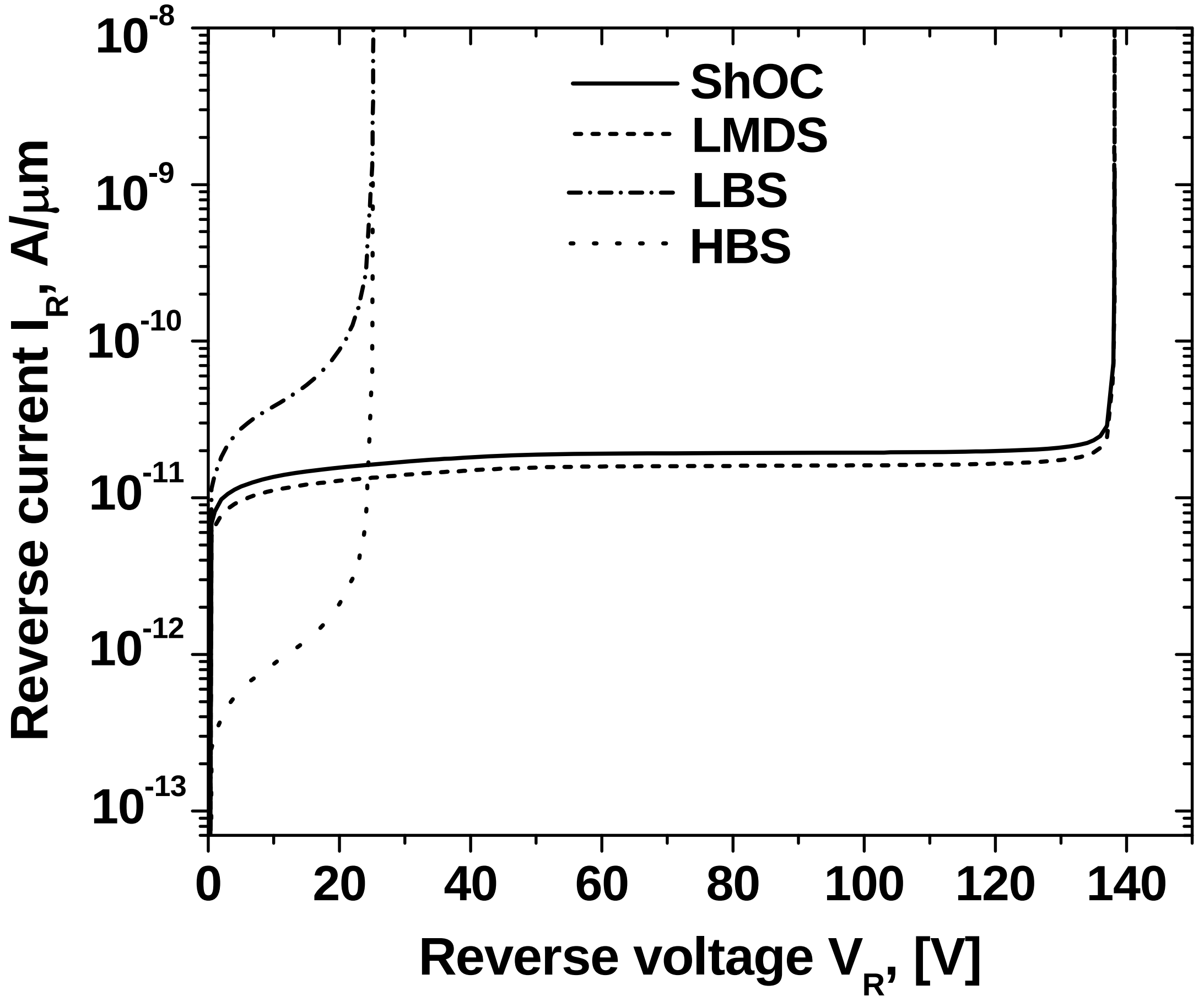

Fig. 2(b)

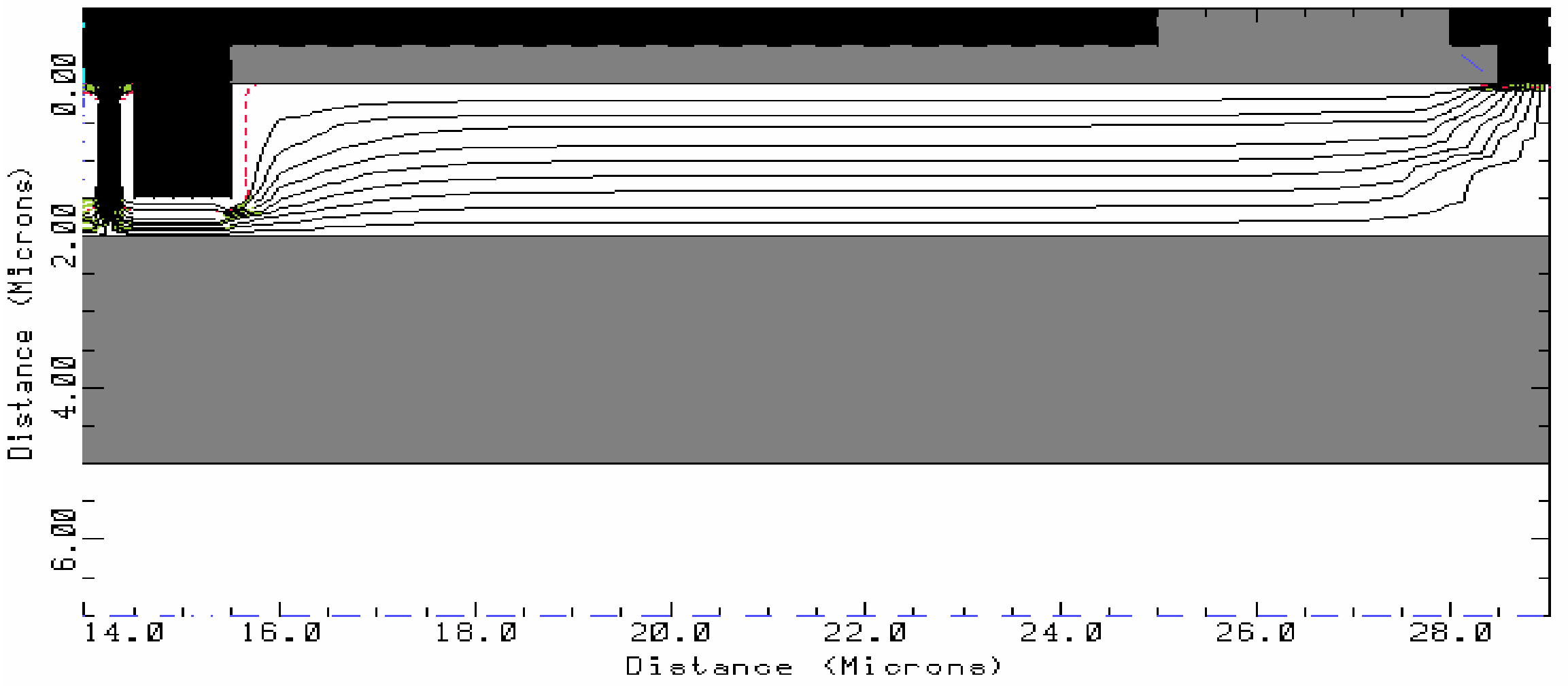

Fig. 3(a)

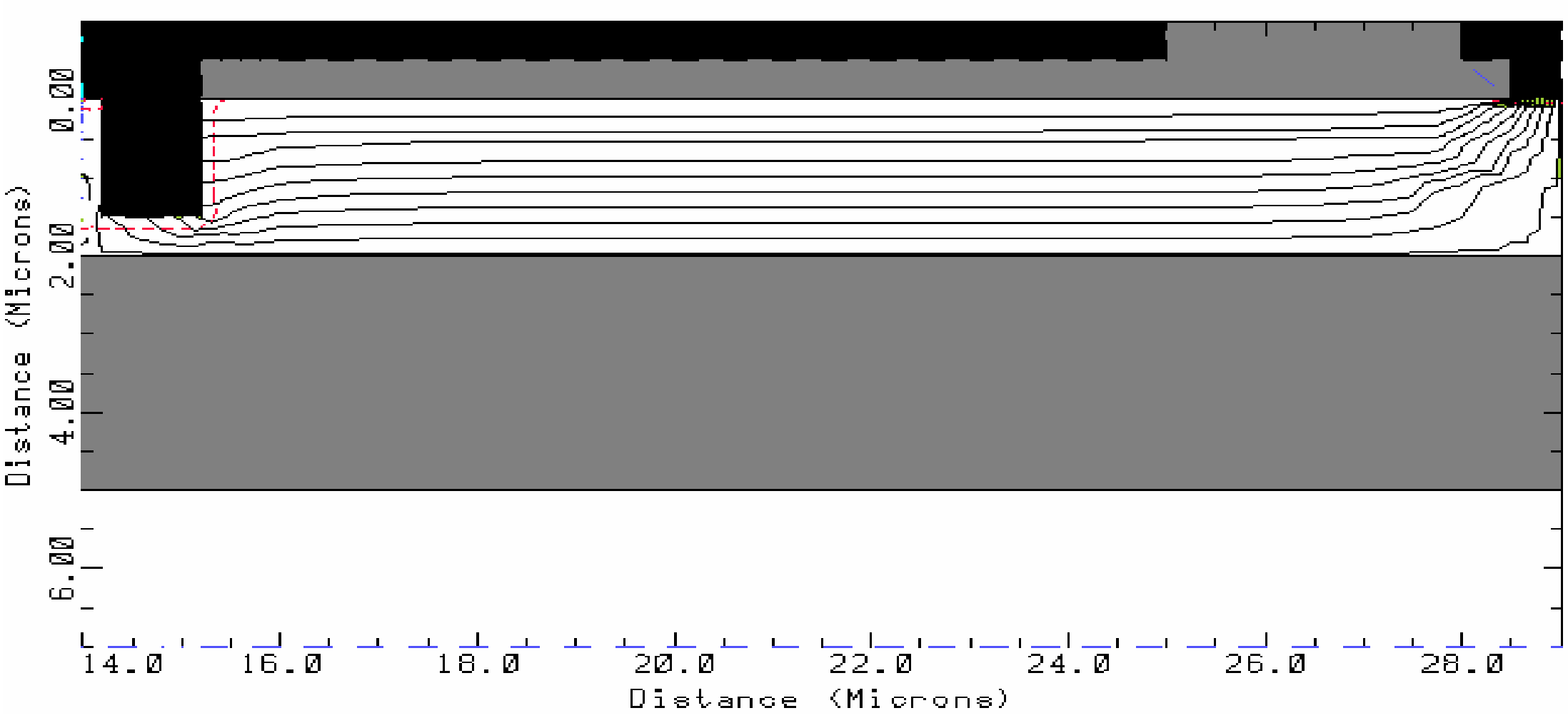

Fig. 3(b)

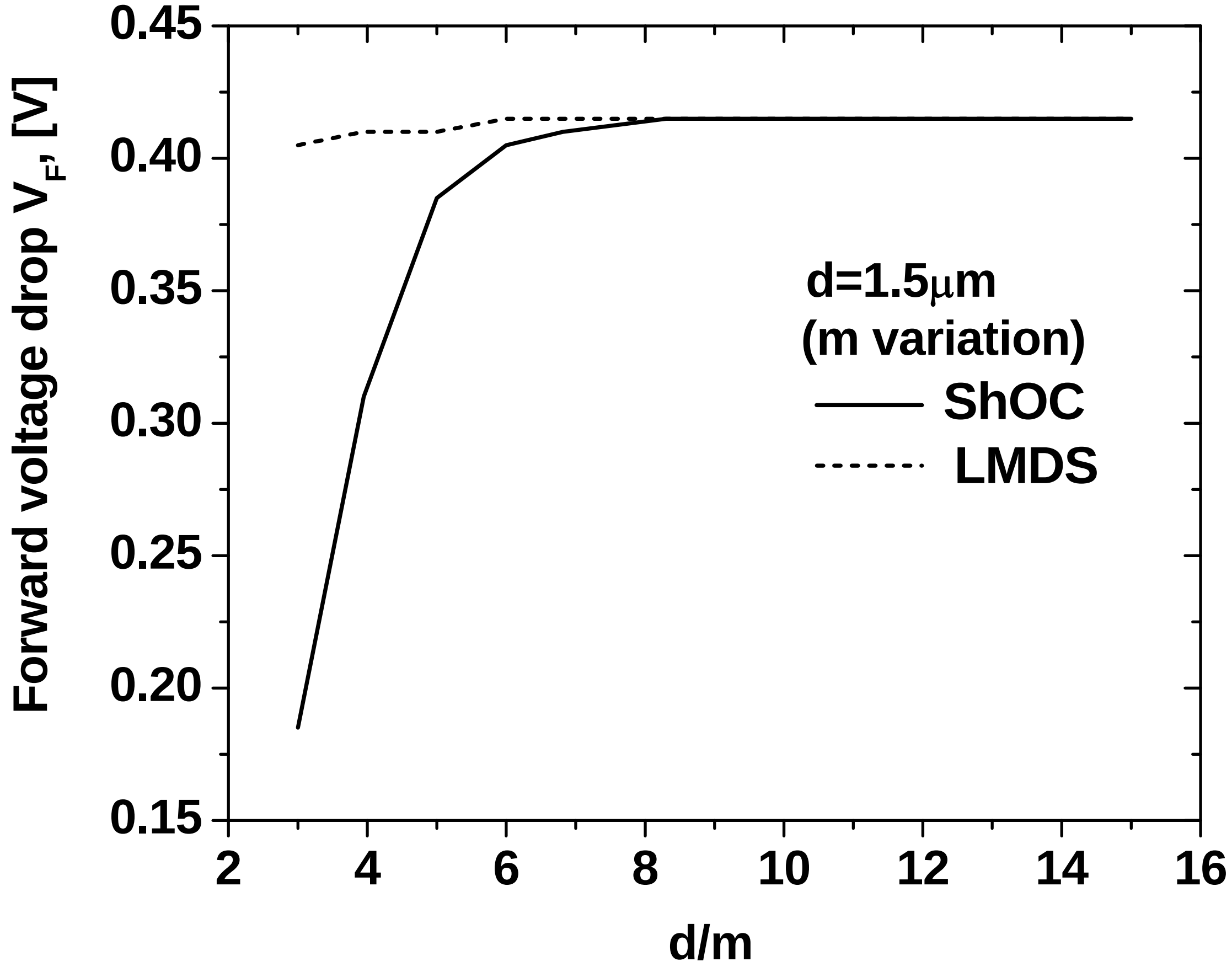

Fig.4(a)

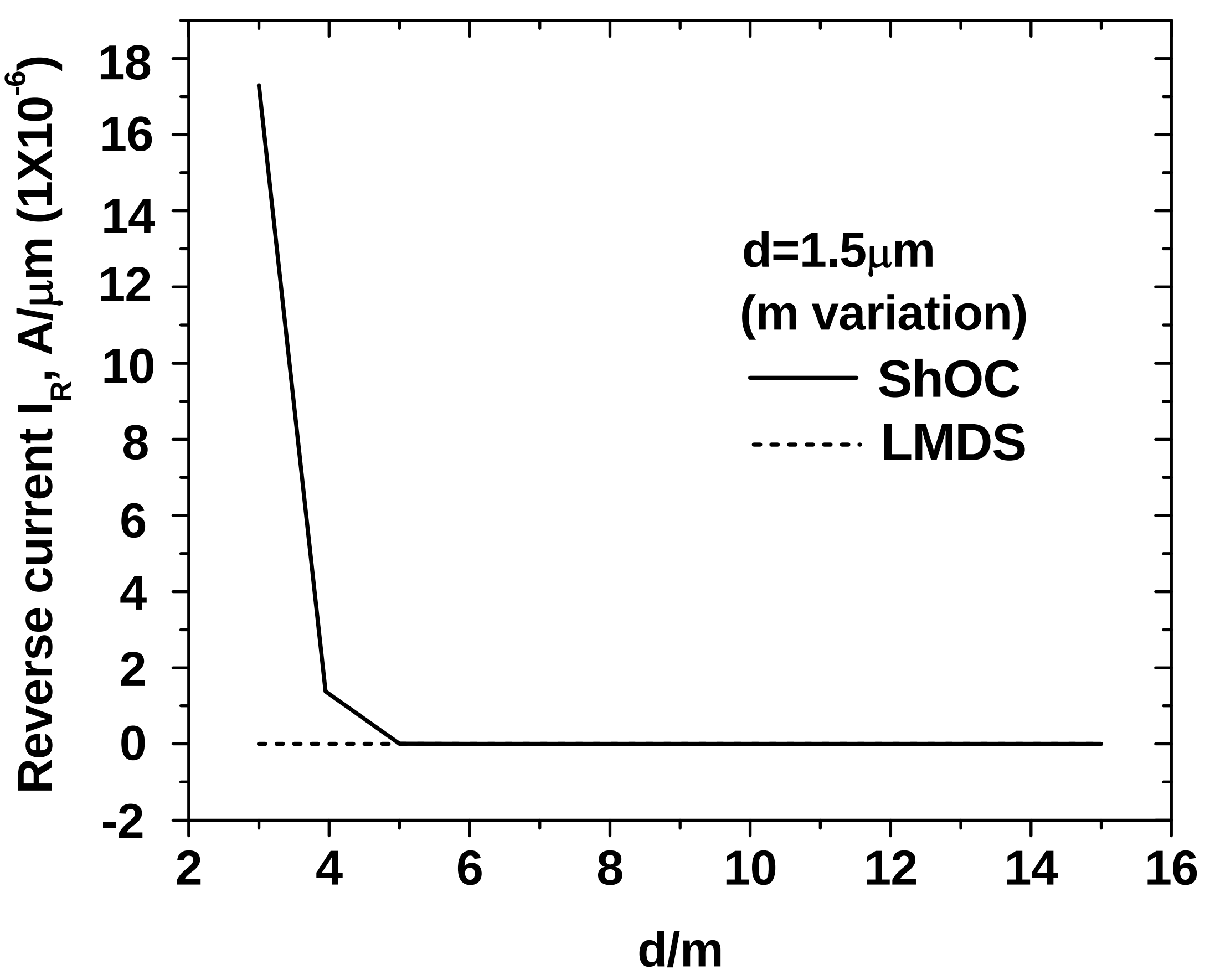

Fig.4 (b)

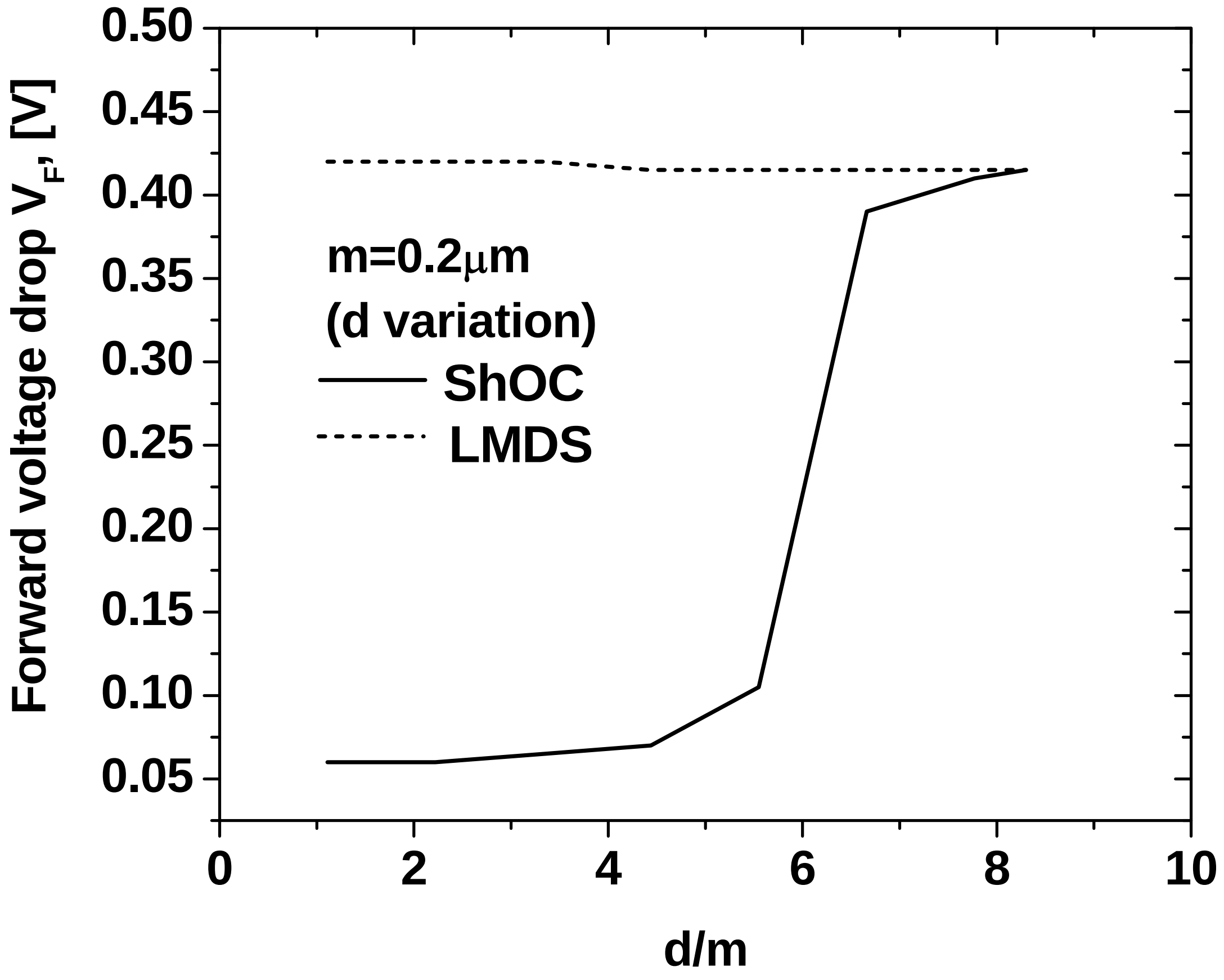

Fig.5 (a)

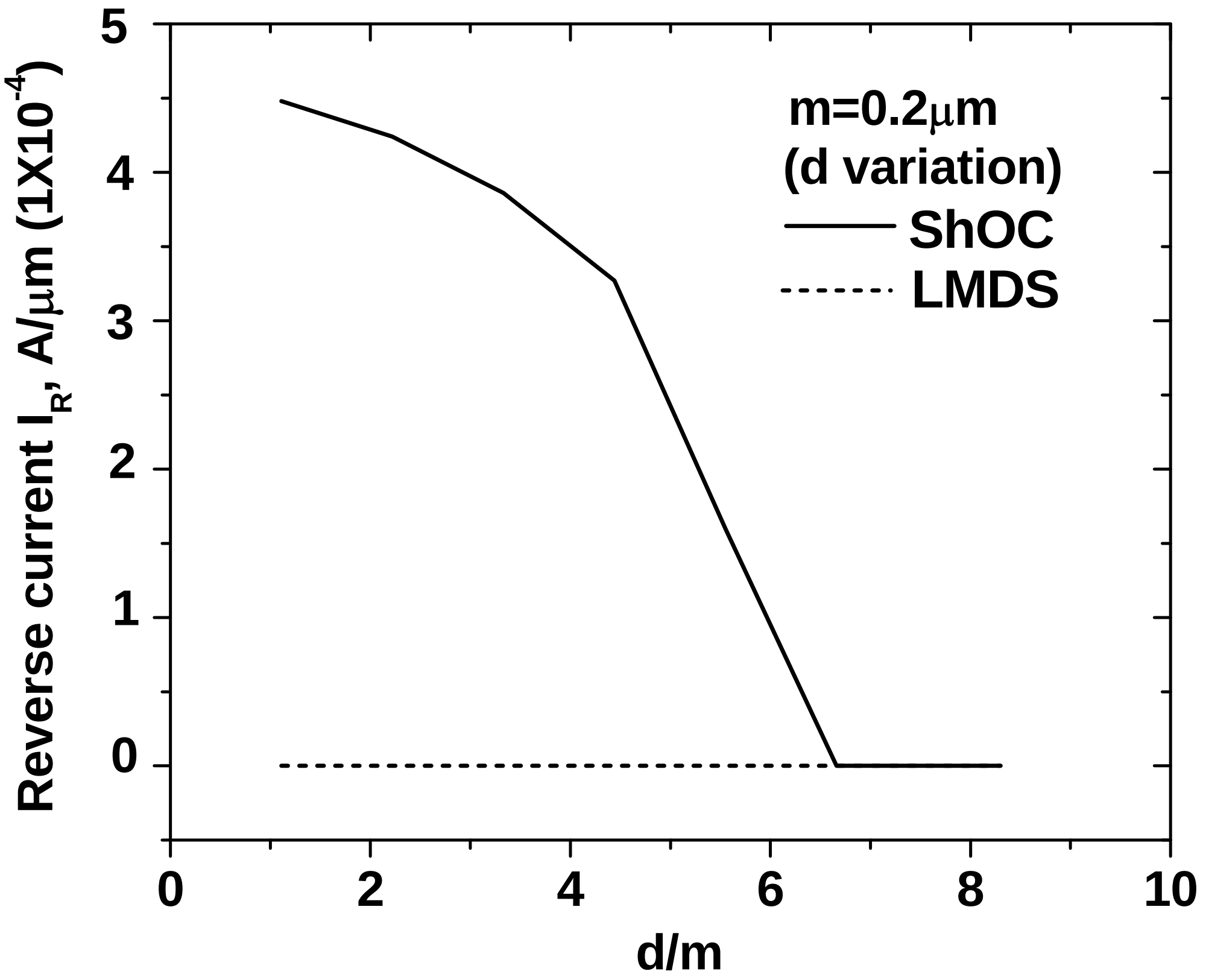

Fig. 5 (b)